\documentclass[a4paper,10pt,conference]{IEEEtran}
\IEEEoverridecommandlockouts

\usepackage{amsmath,amsfonts,amssymb,amsthm}
\usepackage{cite}
\usepackage{array}
\usepackage{graphicx}
\usepackage{mathtools}
\usepackage{graphicx}
\usepackage{textcomp}
\usepackage[dvipsnames]{xcolor}
\usepackage{soul}
\usepackage{float}
\usepackage{stfloats}
\usepackage[caption=false,font=footnotesize,labelfont=sf,textfont=sf]{subfig}
\usepackage{tikz}
\usetikzlibrary{positioning, shapes.geometric, arrows, backgrounds, decorations.pathreplacing}

\usepackage{hyperref}
\hypersetup{
    colorlinks=false,
    hidelinks,
    pdftitle={Towards Bridging the Gap between Near and Far Field Characterizations of the Wireless Channel.},
    pdfauthor={Navneet Agrawal, Ehsan Tohidi, Renato L.G. Cavalcante, Slawomir Stanczak},
    pdfkeywords={Near-field channels, channel modeling, MIMO, Extra-large antennas}
    }
\usepackage{textcomp}

\usepackage[linesnumbered,ruled,vlined]{algorithm2e}

\SetKwInput{KwInput}{Input}
\SetKwInput{KwOutput}{Output}
\SetKwInput{KwInit}{Initialize}
\SetKwInput{KwUpdate}{Update}

\usepackage{pgfplots}
\usepgfplotslibrary{colorbrewer,patchplots}
\pgfplotsset{compat=1.14}
\pgfplotsset{
    discard if/.style 2 args={
        x filter/.code={
            \edef\tempa{\thisrow{#1}}
            \edef\tempb{#2}
            \ifx\tempa\tempb
                
            \fi
        }
    },
    discard if not/.style 2 args={
        x filter/.code={
            \edef\tempa{\thisrow{#1}}
            \edef\tempb{#2}
            \ifx\tempa\tempb
            \else
                
            \fi
        }
    }
}

\def\BibTeX{{\rm B\kern-.05em{\sc i\kern-.025em b}\kern-.08em
    T\kern-.1667em\lower.7ex\hbox{E}\kern-.125emX}}

\title{Towards Bridging the Gap between \\ Near and Far-Field Characterizations of \\ the Wireless Channel\\
\thanks{The authors acknowledge the financial support by the Federal Ministry of Education and Research of Germany in the program of ``Souver\"an. Digital. Vernetzt.'' Joint project 6G-RIC, project identification numbers: 16KISK020K and 16KISK030, and the joint project 6G-LICRIS, project identification numbers: 16KISK141 and 16KISK143.}
}

\author{\IEEEauthorblockN{Navneet Agrawal\IEEEauthorrefmark{1}\IEEEauthorrefmark{2}, Ehsan Tohidi\IEEEauthorrefmark{1}\IEEEauthorrefmark{2}, Renato L.~G.~Cavalcante\IEEEauthorrefmark{2}, S{\l}awomir Sta\'nczak\IEEEauthorrefmark{1}\IEEEauthorrefmark{2}}
\IEEEauthorblockA{\IEEEauthorrefmark{1}Technische Universit\"at Berlin, Germany}
\IEEEauthorblockA{\IEEEauthorrefmark{2}Fraunhofer Heinrich-Hertz-Institut (HHI), Germany}
}

\IEEEpubid{%
    \begin{minipage}{\columnwidth}
    \vspace{1em}
    \copyright 2024 IEEE. Personal use of this material is permitted. Permission from IEEE must be obtained for all other uses, in any current or future media, including reprinting / republishing this material for advertising or promotional purposes, creating new collective works, for resale or redistribution to servers or lists, or reuse of any copyrighted component of this work in other works.
    \end{minipage}%
    \hspace{1em}%
    \begin{minipage}{\columnwidth}
        $ $
    \end{minipage}
}

\DeclareMathOperator*\argmin{arg \, min}		

\newcommand{\Field}[1]{\mathbb{#1}}

\newcommand{\Space}[1]{\mathcal{#1}}
\newcommand{\Set}[1]{\mathcal{#1}}

\newcommand{\Matrix}[1]{\mathbf{#1}}
\newcommand{\Vector}[1]{\pmb{#1}}
\newcommand{\Operator}[1]{\mathsf{#1}}
\newcommand{\Kernel}[1]{\mathtt{#1}}
\newcommand{\Funct}[1]{{#1}}

\newcommand{\abs}[1]{\left|#1\right|}
\newcommand{\norm}[1]{\lVert#1\rVert}

\newcommand{\innerprod}[3]{\left\langle#1, #2\right\rangle_{#3}}

\newcommand{\deriv}[1]{\mathrm{d}#1}

\newcommand{\ji}{\mathrm{j}}

\newcommand{\mylabel}[2]{#2\def\@currentlabel{#2}\label{#1}}

\theoremstyle{remark}
\newtheorem{remark}{Remark}

\newcommand{\Real}{\Field{R}}
\newcommand{\Complex}{\Field{C}}
\newcommand{\Natural}{\Field{N}}

\newcommand{\HS}{\Space{H}}
\newcommand{\HSt}{\HS_T}
\newcommand{\inT}[2]{\innerprod{#1}{#2}{\HSt}}
\newcommand{\normT}[1]{\norm{#1}_{\HSt}}
\newcommand{\HSr}{\HS_R}

\newcommand{\opA}{\Operator{A}}

\newcommand{\kGf}{\Kernel{G}}
\newcommand{\kG}{\kGf_0}

\newcommand{\fE}{\Funct{E}}
\newcommand{\fJ}{\Funct{J}}

\newcommand{\fFYY}{\kGf}

\newcommand{\sC}{\Set{C}}
\newcommand{\sE}{\Set{E}}
\newcommand{\sG}{\Set{G}}

\newcommand{\sT}{\Set{S}_T}
\newcommand{\sTn}{\sT^{(n)}}
\newcommand{\sR}{\Set{S}_R}
\newcommand{\sS}{\Set{S}}
\newcommand{\sU}{\Set{U}}
\newcommand{\sV}{\Set{V}}
\newcommand{\stV}{\tilde{\sV}}

\newcommand{\vg}{\Vector{g}}
\newcommand{\hg}{\hat{\vg}}
\newcommand{\vJ}{\Vector{J}}
\newcommand{\vr}{\Vector{r}}
\newcommand{\tr}{\Tilde{\vr}}
\newcommand{\vs}{\Vector{s}}

\newcommand{\vx}{\Vector{x}}
\newcommand{\vy}{\Vector{y}}

\newcommand{\mI}{\Matrix{I}}

\newcommand{\mM}{\Matrix{M}}

\newcommand{\rs}{r_{s}}
\newcommand{\hr}{\hat{\vr}_s}
\newcommand{\zr}{\hat{\vr}}
\newcommand{\NX}{{N_x}}
\newcommand{\NY}{{N_y}}
\newcommand{\NN}{{N}}

\newcommand{\SIR}{\text{SIR}}
\newcommand{\conC}{\mathfrak{C}}
\newcommand{\fNear}{\kGf_1}
\newcommand{\fFar}{\kGf_2}
\newcommand{\gn}{g^{(1)}}
\newcommand{\gf}{g^{(2)}}
\newcommand{\gi}{g}
\newcommand{\hgi}{\hg}

\newcommand{\vgi}{\vg}

\newcommand{\PEi}{P_E}
\newcommand{\tPE}{\Tilde{P}_E}

\begin{document}

\maketitle

\IEEEpubidadjcol

\begin{abstract}
The ``near-field'' propagation modeling of wireless channels is necessary to support sixth-generation (6G) technologies, such as intelligent reflecting surface (IRS), that are enabled by large aperture antennas and higher frequency carriers.
As the conventional far-field model proves inadequate in this context, there is a pressing need to explore and bridge the gap between near and far-field propagation models.
Although far-field models are simple and provide computationally efficient solutions for many practical applications, near-field models provide the most accurate representation of wireless channels.
This paper builds upon the foundations of electromagnetic wave propagation theory to derive near and far-field models as approximations of the Green's function (Maxwell's equations).
We characterize the near and far-field models both theoretically and with the help of simulations in a line-of-sight (LOS)-only scenario.
In particular, for two key applications in multiantenna systems, namely, beamforming and multiple-access, we showcase the advantages of using the near-field model over the far-field, and present a novel scheduling scheme for multiple-access in the near-field regime.
Our findings offer insights into the challenge of incorporating near-field models in practical wireless systems, fostering enhanced performance in future communication technologies.
\end{abstract}


\section{Introduction}
The rapid evolution of wireless communication technologies has brought about transformative solutions to meet the demands of future applications with high-quality service requirements. 
The increasing need for higher data rates, enhanced energy efficiency, reduced latency, and high reliability has given rise to novel trends in wireless communication development. 
These trends are further driven by a growing focus on campus networks in 6G technologies. 
A reevaluation of wireless channel characterization to ensure reliable and efficient communication is necessitated by 
(i) adoption of large aperture antennas, 
(ii) migration to higher frequency bands, and 
(iii) exploration of new use-cases pertaining to indoor (short-distance) applications. 
Amidst these advancements, the significance of a more accurate channel model has been underscored~\cite{mingyao2023,yuan2022spatial}. 
IRSs exemplify a prominent application area where accurate wave propagation models plays a vital role~\cite{tohidi2023near}.


\begin{figure}[t]
    \centering
    \includegraphics[width=0.9\columnwidth]{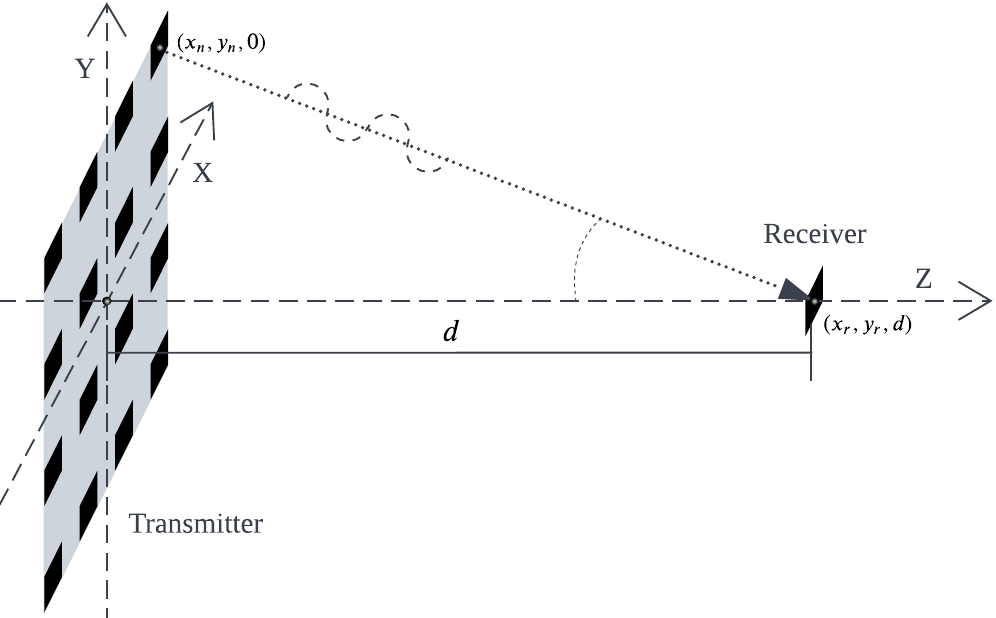}
    \caption{System model}
    \label{fig:system}
\end{figure}

Traditionally, the wireless channel has been characterized separately in the near-field and the far-field.
The near-field corresponds to the regime where the receiver (Rx) is only a few carrier wavelengths away from the transmitter (Tx), and in the far-field the Tx-Rx distance is large enough that the wavefront can be approximated as planar~\cite{pollock2003introducing,abhayapala2003}.\footnote{The study of the region extremely close to the antenna, namely the reactive near-field, is out of the scope of this paper.}~%
In the far-field region, a simplified model is obtained as properties of the electromagnetic (EM) waves stabilize after traveling a sufficient distance.
On the other hand, in the near-field, a more accurate model is required as interactions with the environment are more pronounced.
The boundary between near and far-field behaviors is traditionally demarcated via the Fraunhofer or Rayleigh distance, but these estimates are often very conservative~\cite{bjornson2020power}. 
As the relevance of the near-field grows, alongside the established importance of far-field characterizations, there is an increasing need to bridge the gap between these two characterizations.

Recently, extensive research has emerged in the area of wireless communication exploring various directions and perspectives that distinguish between near and far-field models~\cite{liu2023near,ramezani2022near,dardari2020communicating,bjornson2020power,yu2023near,yijin2023RIS,zhang20226g,wei2022codebook}. 
For instance, authors in \cite{liu2023near} offer a comprehensive review of near-field communications, encompassing channel modeling, beam focusing, antenna architectures, and performance analysis. 
In \cite{yu2023near}, channel estimation for mixed LOS/NLOS environments in extremely large-scale multiple-input multiple-output (MIMO) systems within the near-field regime is presented. 
Moreover, IRS-aided localization and channel estimation for Terahertz systems, exploring the near-field model, are proposed in \cite{yijin2023RIS}. 
The Author in \cite{dardari2020communicating} delves into the near-field model for holographic antennas, which offer increased control and degrees of freedom in the near-field, but pose challenges in terms of implementation and cost. 
Other studies, such as \cite{ramezani2022near,bjornson2020power}, take a pragmatic systems approach, focusing on how specific metrics like received power or beam width change in relation to the chosen model with the distance between the Tx and the Rx.

This paper focuses on bridging near and far-field models of the EM wave propagation for wireless communication using Maxwell's equations. 
We revisit the fundamentals behind developing the near-field models from EM wave theory, and incorporate these models in various applications, with the aim of helping researchers and engineers in understanding the foundations.
We strive to elucidate the analogy between the standard far-field MIMO model and the proposed models derived from Green's function by designing a pragmatic MIMO system accordingly. 
While Green's function, derived from solving Maxwell's equations, provides an accurate representation, its relation to the existing far-field models is often unclear in the literature.
In \cite{poon2005degrees}, the multipath channel in the near-field is characterized using a simpler model that consists of a cascade of linear operators, motivated by the similar approach used previously in the far-field models.
Our approach in this paper is to first elucidate how Maxwell's equations represent the wireless channels, and lead to the typical linear channel model used in the traditional literature \cite{tse2005fundamentals}.
Starting from the Green's function, we successively derive the near-field model, where the wavefront is still spherical, and the far-field model, which results from the planar-wave approximation.
Furthermore, we use the proposed models to extend two pivotal applications in the MIMO systems, namely, beamforming and multiple-access, from far-field perspective to near-field. 
These applications clarify the gap between near and far-field characterizations of the channel for different regions of the propagation environment. 
In particular, in the near-field, we show that a more accurate channel model can lead to new use cases for the applications, which are not possible using the far-field models.

\setcounter{equation}{1}

\begin{figure*}[b!]
\begin{equation}
    \label{eq:Greens_func}
    \begin{aligned}
        (\forall \vx\in\Real^3, \norm{\vx}\neq 0), \quad 
            \kG(\vx) &= -\frac{\ji \eta e^{-\ji \frac{2\pi}{\lambda} \norm{\vx}}}{2 \lambda} \Bigg[ \frac{1}{\norm{\vx}}\left(\mI - \frac{\vx\vx^T}{\norm{\vx}^2} \right) + \frac{\ji \lambda}{2\pi \norm{\vx}^2} \left(\mI - 3\frac{\vx\vx^T}{\norm{\vx}^2} \right) + \frac{-\lambda^2}{4\pi^2 \norm{\vx}^3} \left(\mI - 3\frac{\vx\vx^T}{\norm{\vx}^2} \right) \Bigg]
    \end{aligned}
\end{equation}
\end{figure*}

\section{A Gentle Summary of Wireless Channel Modeling} \label{sec:models}
In this section, we summarize some of the recent results from~\cite{Miller2019,bjornson2020power,dardari2020communicating,ramezani2022near,mingyao2023,zhang20226g}.
Our contribution here lies in providing analogies between standard channel modeling approaches and those derived from Green's function.
The aim is to provide engineers and researchers with novel insights that may help them to apply techniques designed using standard far-field models to new analyses and applications pertaining to the near-field models.
In this study, we focus on a scattering-free environment, i.e., communication occurring exclusively through the line-of-sight (LOS) link.
This system allows for deterministic modeling of the wireless channel using Maxwell's equations, providing a solid theoretical foundation for our study. 
While the LOS-only scenario may seem limiting, it finds practical relevance in systems utilizing high-frequency carriers for indoor communication.

The system configuration comprises a Tx, with its antenna surface placed along the XY-plane and centered at the origin, and an Rx, with its antenna surface placed parallel to the XY-plane and centered on the YZ-plane.
An illustration of the system is provided in Fig.~\ref{fig:system}.
We consider an isotropic uniform medium, such as air, with constant scalar permeability ($\mu$), and permittivity ($\epsilon$).
Additionally, we restrict our analysis to monochromatic fields, i.e., fields whose dependence on time $t$ takes the form $\sin(2\pi f t)$, $\cos(2\pi f t)$, $\exp(\ji 2\pi f t)$, or $\exp(-\ji 2\pi f t)$, or linear combinations of these, with a specific choice of frequency $f$, where $\ji$ denotes the unitary imaginary number (i.e.,~$\ji^2 = -1$).
These environment assumptions enable closed-form solutions of Maxwell's equations without limiting the general modeling of the wireless channel (see \cite{Miller2019}).

In essence, the Tx conveys information to the Rx over the wireless medium as follows: 
The Tx antenna is excited with a current carrying the transmitted information, which in turn induces an electric field in its vicinity. 
In response, the Rx antenna captures the electric field induced on its surface and decodes the information it carries. 
The relation between the excited current at the Tx and the induced electric field at a point in its vicinity is governed by Maxwell's equations.
For this system, solving Maxwell's equations lead to a linear operator mapping input current to an induced electric field, described by the \emph{Green's function} \cite{poon2005degrees,Miller2019,dardari2020communicating}.

In the following, we present the closed-form expression of Green's function and introduce two approximations of Green's function, which are contingent on the distance between the Tx and the Rx: one in the near-field and the other in the far-field.
In Section \ref{sec:system}, we apply these prescribed models to a pragmatic multiple-input single-output (MISO) system.

\setcounter{equation}{0}

\subsection{Greens' function and its approximations} \label{sec:Green}
According to Maxwell's equations, the electric field induced at a point $\vr\in\Real^3$ due to the current density $\fJ:\sT\to\Complex^3$ excited on the Tx antenna surface $\sT \subset \Real^3$, is given by \cite{poon2005degrees,Miller2019,dardari2020communicating}:
\begin{align}
    \label{eq:wave_eq}
    (\forall \vr\in\Real^3 \setminus \sT) \qquad \fE(\vr) = \int_{\sT} \kG(\vr, \vs) \fJ(\vs) \deriv\vs \,\, \in\Complex^3,
\end{align}
where $\kG:\Real^3\times\Real^3\to\Complex^{3\times 3}$ is the Green's function.
The closed-form expression of the Green's function is given in \eqref{eq:Greens_func} at the bottom of this page, which shows that $\kG$ is in fact a shift-invariant kernel, i.e., it only depends on the vector $\vx := \vr - \vs$, the position of the Rx w.r.t.~the Tx.
The three output dimensions of $\fJ$ and $\fE$ represent the EM wave polarization in X, Y, and Z directions, respectively.
In \eqref{eq:Greens_func}, and for the rest of this paper, we use standard $2$-norm on vectors in $\Real^3$ given by $(\forall\vx\in\Real^3)\ \norm{\vx} := (\sum_{i=1}^3 \vx_i^2)^{\frac{1}{2}}$.
Furthermore, the symbols $\eta = \sqrt{\mu/\epsilon}$ and $\lambda$ denote the impedance of free space and the wavelength, respectively.

In the following, as in \cite{dardari2020communicating}, we assume that the current density $\fJ$ is an element of the Hilbert space $\HSt$, defined as $\HSt:= \{f:\sT \to \Complex^3:\normT{f}^2 < \infty\}$ with inner product $\inT{f}{g} := \int_{\vs} f^H(\vs) g(\vs) \deriv\vs$, and induced norm $\normT{f} = \sqrt{\inT{f}{f}}$.
Let the Rx antenna surface be $\sR \subset \Real^3$, and assume that the electric field $\fE:\sR\to\Complex^3$ is an element of the Hilbert space $\HSr$, defined analogously to $\HSt$.
Then, equation \eqref{eq:wave_eq} can be represented via a bounded linear operator $\opA:\HSt\to\HSr$, given by $\opA(\fJ)(\vr) := \int_{\sT} \kG(\vr - \vs) \fJ(\vs)$, such that \eqref{eq:wave_eq} becomes $\fE = \opA(\fJ)$.

\setcounter{equation}{2}

\begin{remark}[Linear model of the wireless channel]
    In standard wireless communication literature \cite[Ch.~2]{tse2005fundamentals}, the channel is consistently modeled as a linear system, and the validity of this approach is demonstrated by \eqref{eq:wave_eq}.
    In essence, \eqref{eq:wave_eq} expresses that the total electric field at $\vr$ results from a weighted summation of $\fJ(\vs)$ for each $\vs\in\sT$, with weights given by $\kG(\vr, \vs)$, which represents the wireless channel between points $\vs\in\sT$ and $\vr\in\sR$.
    The current density and electric field can be embedded into Hilbert spaces of functions, respectively, and then, \eqref{eq:wave_eq} becomes $\fE = \opA(\fJ)$, where $\opA$ is a bounded linear mapping, expressing the wireless channel transfer function.
    The linear operator $\opA$ is modeled using the near or the far-field models, and its properties, such as eigen-functions \cite{dardari2020communicating}, enables many prominent applications in modern wireless communication systems.
\end{remark}

The Green's function is a complicated function, but fortunately its simplified approximations are often enough for modeling the channel.
Note that the second and third terms in \eqref{eq:Greens_func} are proportional to the inverse of the distance between Tx and Rx, i.e., $\norm{\vx}^{-2}$ and $\norm{\vx}^{-3}$, respectively, and hence, they quickly become insignificant as the distance grows.
The first approximation of \eqref{eq:wave_eq} is obtained by removing these terms from the Green's function.
We also rearrange the expression of $\kG$ by separating the Tx antenna position $\vs\in\sT$ dependent terms from those independent of it.
For clarity, we use the following shorthand notation: $\rs := \norm{\vr - \vs}$, $\hr := (\vr - \vs) / \rs$, $r := \norm{\vr}$, $\zr := \vr/\norm{\vr}$, $\omega := \exp(-\ji 2\pi \lambda^{-1} r)$ and $\mM_1(\hr) := (\mI - \hr\hr^T)$, where $\mI$ denotes the identity matrix.
The first approximation of $\kG$, which leads to the near-field model of the wireless channels, is given by:
\begin{align}\label{eq:fNear}
    \fNear(\vr, \vs) := -\frac{\ji \eta \omega}{2 \lambda r}\ \left[e^{-\ji \frac{2\pi}{\lambda} (\rs - r)} \frac{r}{\rs}\ \mM_1(\hr)\right].
\end{align}



The second approximation leads to the standard far-field model with the approximation that the distance of any Tx antenna from the Tx center (the origin) is much less than the distance of the Rx, i.e., $\norm{\vs} \ll \norm{\vr}$.
The far-field approximation, given below in \eqref{eq:fFar}, follows from \eqref{eq:fNear} by using the first-order Taylor approximation of the multivariate function $f(\vs) := \norm{\vr - \vs}$ at $\vs=(0,0,0)$ for a given $\vr$, which gives $\norm{\vr - \vs} \approx \norm{\vr} - \zr^T\vs$, where $\zr := \vr/\norm{\vr}$.
Thus, the far-field approxmiate model is given by:
\begin{align}\label{eq:fFar}
    \fFar(\vr, \vs) := -\frac{\ji \eta \omega}{2 \lambda r}\ \mM_1(\zr) \left[e^{\ji \frac{2\pi}{\lambda} (\zr^T\vs)}\right].
\end{align}
Note that only the term inside the square brackets depends on the position of the Tx antenna element $\vs$, while other only depend on the relative position of the Rx w.r.t.~the Tx center.

The models $\kGf_i$, for $i=0,1,2$, define the behavior of the EM waves in a point-to-point propagation.
In the next section, we apply these models to a pragmatic system that accounts for the practical limitations of the existing technologies.

\subsection{Pragmatic system design} \label{sec:system}

As depicted in Fig.~\ref{fig:system}, our system involves a MISO configuration with a rectangular grid of $\NN := \NX \NY$ (square-shaped) small antenna elements at the Tx.
The antenna elements are indexed as $n=1, \dots, N$, uniformly arranged on the XY-plane, with $\NX$ and $\NY$ antennas along the X and Y axes, respectively. 
Any antenna element $n$, with surface denoted by $\sTn$, has a small area $a^2$, $a\leq 0.5\lambda$, and it is centered at $\vs_n$. 
We assume that all antenna elements are identical and can transmit or receive waves with only Y-polarization. 
Additionally, each antenna element is fed with a constant current density over its surface, i.e., $(\forall \vs \in\sTn)\ \fJ(\vs) = J_n := \fJ(\vs_n)$. 
The spacing between any two adjacent antenna elements along the X-~or the Y-axis is set to $0.5\lambda$.
The Rx consists of a single antenna element, identical to the antenna elements used in the Tx.
For the system described above, \eqref{eq:wave_eq} can be written as \cite{ramezani2022near}:
\begin{align}
    \label{eq:b_E}
    \fE(\vr) = \sum_{n=1}^N J_n \int_{\vs\in\sTn} \fFYY(\vr, \vs)\ \deriv \vs =: \sum_{n=1}^N J_n\ \gi_n(\vr),
\end{align}
where $\fFYY:\Real^3\times\Real^3\to\Complex$ denotes the kernel corresponding to $i=0,1,2$, i.e.,~$\fFYY$ is either the Green's function $\kG$, its near-field approximation $\fNear$, or far-field approximation $\fFar$.
Notice that $\fFYY$ outputs a complex scalar, corresponding to the middle row and column component of $\Complex^{3\times 3}$ the original Green's function output.
This is because, by design, only the Y-polarization components of the Tx current $\fJ$ and the Rx electric field $\fE$ are nonzero.
We define $\gi_n(\vr) := \int_{\vs\in\sTn} \fFYY(\vr, \vs) \deriv \vs$ as the total influence of the $n$th Tx antenna element on the electric field at $\vr$.

Since individual antenna elements are small, the electric field generated by any point on the surface $\sTn$ of $n$th antenna element is approximately the same.
With this approximation, the function $\gi_n(\vr)$ for the near and far-field models are given by: 
\begin{align*}
    \gn_n(\vr) &= a^2 P_1(\vr) \left[e^{-\ji \frac{2\pi}{\lambda} (r_n - r)} \frac{r}{r_n} \mM_1(\hat{\vr}_n) \right], \\
    \gf_n(\vr) &= a^2 P_1(\vr) \mM_1(\zr)\ \left[ e^{j \frac{2\pi}{\lambda} (\zr^T \vs_n)} \right],
\end{align*}
where $r_n:=\norm{\vr - \vs_n}$, $\hat{\vr}_n := (\vr - \vs_n)/r_n$, and $P_1(\vr) := -\frac{\ji \eta \omega}{2 \lambda r}$ collects the path loss and the phase shift that only depends on $\vr$.

\begin{remark}[Analogies with the standard far-field models]
    Notice that, in the far-field model $\gf_n(\vr)$ above, the Tx antenna element position $\vs_n$ appears solely in the phase rotation term $e^{\ji \frac{2\pi}{\lambda} (\zr^T \vs_n)}$.
    In the standard far-field models, these phase terms are typically collected in an array known as the antenna array response.
    However, in the near-field model $\gn_n(\vr)$, an antenna element introduces both phase and amplitude effects, which are now functions of the relative position $(\vr - \vs_n)$ of the Tx antenna element.
    Consequently, the near-field model requires additional information, specifically the precise positions of both the Tx and Rx, in contrast to the far-field model, which relies solely on their angular orientation.
\end{remark}

\section{Applications}

\begin{figure*}[t]
    \includegraphics[width=\textwidth]{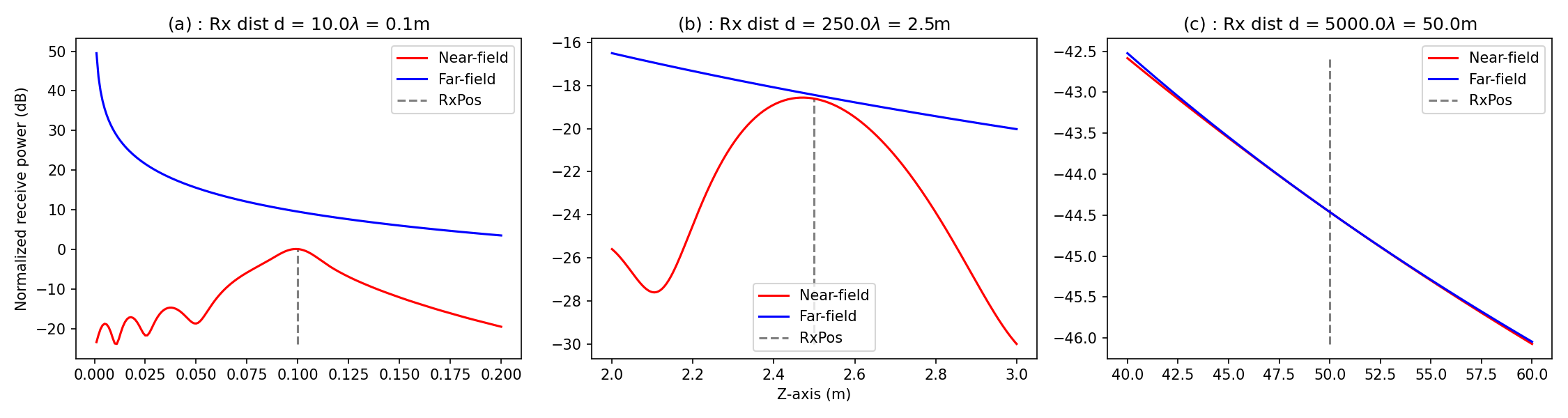}
    \caption{Comparison of near and far-field models for different Rx distances (Z-axis) from Tx.}
    \label{fig:comparison}
\end{figure*}

In this section, we compare two important applications of MIMO systems, namely beamforming and multiple access, using near and far-field modeling approaches.
We show that the adoption of the near-field model not only enhances performance but also unlocks new degrees of freedom beyond the confines of the traditional far-field model. 
By exploring these applications, we will underscore the imperative need for accurate channel models that account for the near-field effects.

The Rx, with antenna surface $\sR$ of area $a^2$ and centered at $\vr$, is assumed to be small enough so that it receives approximately the same electric field $\fE(\vr)$ at any point on its surface.
The power of received electric field at the Rx via \eqref{eq:b_E} is given by \cite{dardari2020communicating,ramezani2022near}:
\begin{align} \label{eq:b_P}
    P(\vr) = \abs{\int_{\sR} \fE(\tr) \deriv \tr}^2 = a^2 \abs{\fE(\vr)}^2 = a^2\ \abs{\sum_n \fJ_n \gi_n(\vr)}^2.
\end{align}

\subsection{Beamforming by spatial focusing} \label{sec:beamformer}
The objective of the beamforming application considered here is to find the current excitation $\fJ_n, n=1,\dots,N$, at each Tx antenna element, that maximizes the power $P(\vr)$ at the Rx position $\vr$ under the transmit power constraint $\sum_n \abs{\fJ_n}^2 = 1$.
To keep the notation uncluttered, we drop the index $i$ (used to specify the model in \eqref{eq:b_E}) in the following.
For the Rx at $\vr$, this is achieved by the matched-filtering (MF) beamformer given by $(\forall n)\ \fJ_n(\vr) = \gi_n^\star(\vr)/\norm{\vgi(\vr)}$, for either $i=1$ or $i=2$, which is also optimal in this case \cite{bjornson2020power}.
Here, $x^\star$ denotes the complex conjugate of $x\in\Complex$, and define $\vgi(\vr) := [\gi_1(\vr), \dots, \gi_N(\vr)]^T$.
When the Tx employs the MF beamformer focusing the beam at the location $\vr$ of the Rx, i.e.,~$\vJ(\vr):= [\fJ_1(\vr), \dots, \fJ_N(\vr)]^T$, then the power of the received electric field at any Rx position $\vr'$ is given by \cite{bjornson2020power}:
\begin{align} \label{eq:MFB_power}
    \PEi(\vr, \vr') := a^2 \norm{\vgi(\vr')}^2\ \abs{(\hgi(\vr))^H \hgi(\vr')}^2,
\end{align}
where, for $\vx\in\Real^3$, we define $\hgi(\vx) := \vgi(\vx)/\norm{\vgi(\vx)}$.
Note that the inner product $\abs{(\hgi(\vr))^H \hgi(\vr')}$ is equal to $1$ if and only if $\hgi(\vr') = \pm \hgi(\vr)$, otherwise, it's smaller than $1$.
In other words, $\PEi(\vr, \vr')$ is maximized at the Rx position $\vr' = \pm \vr$ for the near-field model, and for $\zr' = \pm \zr$ for the far-field model.
The expressions for $\gn_n(\vr)$ and $\gf_n(\vr)$ are provided in Section \ref{sec:system}.

The figure of merit is the \emph{normalized} received power in decibels given by $\tPE(\vr, \vr') := 10\log_{10}(P_E(\vr,\vr')/P_E(\vr_u,\vr_u))$, normalized such that for Rx at $\vr_u = (0, 0, 0.1m)$, we get $\tPE(\vr_u, \vr_u) = 0$dB using the near-field model $\fNear$.
In the simulation, we use a carrier frequency of $30$GHz with wavelength $\lambda=0.01m$, the number of antenna elements along X and Y axes is $\NX=20$ and $\NY=200$, respectively, and antenna element size is $a=0.5\lambda$.
In Fig.~\ref{fig:comparison} (a), (b) and (c), the MF beamformer $\vJ(\vr)$ is designed for maximizing the received power at the Rx located at $\vr=(0,0,d)$ with $d=10\lambda=0.1$m, $d=250\lambda=2.5$m, and $d=5000\lambda=50$m, respectively.
From plots (a) and (b), the discrepancy in the near and far-field models is apparent.
In the radiated near-field scenario, the near-field model allows power to be concentrated at a specific Rx position, and for $d' = d \pm \delta$, $\delta>0$, the power decreases.
In the far-field, in contrast, Tx can only concentrate the power towards an angular orientation of the Rx.
When the distance between the Tx and the Rx is large enough, as in Fig.~\ref{fig:comparison}(c), the near and far-field models almost coincide, confirming that the far-field model $\fFar$ is in fact an approximation of the near-field model $\fNear$.
We remark that for the given Tx antenna configuration, the \emph{Fraunhofer distance} is $\frac{2D^2}{\lambda} \approx 200$m, where $D = 0.5\lambda \sqrt{\NX^2 + \NX^2}$ is the diameter of the Tx antenna.
Clearly, the Fraunhofer distance gives a very conservative estimate of the distance at which far-field models can be reliably employed.
It is evident from this discussion that the near-field model provides a more accurate description of the wireless propagation, and facilitates the use of an additional degree of freedom (the distance $d$) for improving performance in many applications.
Observe that in \ref{fig:comparison}(a) with $d=0.1m$, the far-field model overestimates the received power.
This can be explained as follows:
In the far-field model of \eqref{eq:fFar}, the path-loss is the same for all antenna elements, proportional to $\norm{\vr}^{-1}$, whereas, in the near-field, the path-loss is proportional to $\norm{\vr - \vs_n}^{-1}$.
This leads to a lower total received power in the near-field, as $\norm{\vr} \leq \norm{\vr - \vs_n}$ for all $n$.


\subsection{Multiple Access in Near-Field} \label{sec:multiple_access}

A key application of MIMO systems is enabling multiple users to be served simultaneously such that the received signal obeys a minimum SIR requirement.
In this section, we assume the channel to be deterministic and static.
However, the method presented here can be applied to stochastic time-varying channels using covariance matrix estimates \cite{cavalcante2020channel,agrawal2021aps}.
Consider a scenario with two receivers, henceforth called users, indexed $1$ and $2$, with their channels denoted by $\vg_1$ and $\vg_2$, respectively.
With the prior channel knowledge, the Tx constructs a signal $s(t)$ as $(\forall t\in\Natural)\ s(t) = \hg_1^\star x_1(t) + \hg_2^\star x_2(t)$, where, for user $i=1,2$, $x_i(t)\in\Complex$ is the signal to be transmitted to $i$, and $\hg_i := \vg_i/\norm{\vg_i}$ is the MF beamformer focusing the energy at user $i$ (see Section \ref{sec:beamformer} above).
In response, the signal received by $i$ is $r_i(t) = \norm{\vg_i}\ x_i(t) + \norm{\vg_i} (\hg_i^H\hg_j)\ x_j(t)$, where $j=1,2, j\neq i$.
Note that we ignore the receiver noise here for simplicity.
To ensure that users $1$ and $2$ receive their corresponding signals with a SIR greater than a minimum threshold $\gamma>0$, one way is to find a suitable pair of positions for the two users such that the inner-product $\abs{\hg_1^H\hg_2}^{-2} \geq \gamma$.
In this way, a multi-antenna Tx can provide multiple single-antenna users with separate data streams simultaneously over the same time and frequency resource of the wireless medium.

In a multi-user scenario, the problem is formulated as finding the subset of user locations that can be served simultaneously while satisfying the desired minimum SIR requirement.
In this direction, consider $K$ user positions indexed by $\sU := {1,\dots, K}$, and position them along the same angular orientation w.r.t.~the Tx, i.e.,~placed along a ray $\hr$.
Without loss of generality, we select $\hr = (0, 0, 1)$, i.e.,~users are placed on the Z-axis.
Note that serving these users simultaneously over the same wireless resource is not feasible when utilizing the far-field model to represent the channels, as discussed in Section \ref{sec:beamformer}.

Let the channel between the Tx and user located at a position indexed by $k\in\sU$ is given by the vector $\vg_k := [\fNear(\vr_k, \vs_1), \dots, \fNear(\vr_k, \vs_\NN)]^T$, where $\fNear$ is the near-field channel model in \eqref{eq:fNear}.
Given channels of users $k$ and $l$, where $l\neq k$, the Tx sends the signal $(\forall t\in\Natural)\ s(t) = \hg_k^\star x_k(t) + \hg_l^\star x_l(t)$, where $\abs{x_k(t)} = 1$ and $\abs{x_l(t)} = 1$ for all $t$.
Then, the SIR at the user $l$ (similarly, at user $k$) is given by:
\begin{align} \label{eq:SIR}
(\forall l\in\sU, \forall k\in \sU\setminus \{l\}) \qquad    \SIR(l,k) := \abs{\hg_k^H\hg_l}^{-2}.
\end{align}
Note that the function $\SIR$ is symmetric.
Our objective is to find the largest set $\sV\subseteq\sU$ such that for any distinct pair of user positions $k$ and $l$ in $\sV$, $\SIR(l,k)$ and $\SIR(k,l)$ are greater than $\gamma$.
We remark that the pairwise SIR requirement in the above approach can be easily extended to SIR with multiple interfering users by increasing the SIR threshold $\gamma$ such that the desired SIR is always achieved.


We use a graph-theoretic approach to this problem.
Define a graph $\sG:=(\sU, \sE)$ with $K$ nodes such that an edge exist between any two nodes $k$ and $l$ if the condition $\conC(k, l) := (\SIR(l,k) > \gamma$ {\bf and} $\SIR(k,l) > \gamma)$ is satisfied.
More precisely, $(k,l)\in\sE$ if $\conC(k,l) = \text{True}$.
The problems of finding $\sV \subset \sU$ for which the minimum SIR condition is satisfied can be formulated as the \emph{maximum clique problem} in the (undirected) graph $\sG$ with edge-set $\sE$ \cite{douik2020tutorial}.
It is well-known that the maximum clique problem is \emph{NP-complete} \cite{wu2015review}.
In this paper, we present a heuristic algorithm, given by Algorithm \ref{alg:1}, which finds the set $\stV$, a close approximation of the optimal set $\sV$.
Intuitively, the algorithm iteratively adds the point closest to the Tx from the set $\sU$ to $\stV$, say $k\in\sU$, and then removes points $j\in\sU$ that do not satisfy the SIR condition $\conC(j,k)$.

\begin{algorithm}
\DontPrintSemicolon
\KwInput{$\sU :=$ User (Rx) locations, $\gamma :=$ minimum SIR }
\KwOutput{$\sG \rightarrow$ Set of selected users.}
        $\sS \leftarrow \sU$, $\sG \leftarrow \emptyset$\;
        \While{$\sS \neq \emptyset$}{
            $\vx \leftarrow \argmin_{\vx\in\sS} \norm{\vx}$\;
            $\sG \leftarrow \sG \cup \{\vx\}$\;
            $\sC \leftarrow \{\vy \in \sS \mid \SIR(\vx, \vy) \leq \gamma$ {\bf or} $\SIR(\vy, \vx) \leq \gamma\}$\tcc*{see eq.~\eqref{eq:SIR}}
            $\sS \leftarrow \sS \setminus \sC$\;
            }
\caption{Heuristic Algorithm for User Selection}
\label{alg:1}
\end{algorithm}

\begin{figure}
    \centering
    \includegraphics[width=0.9\columnwidth]{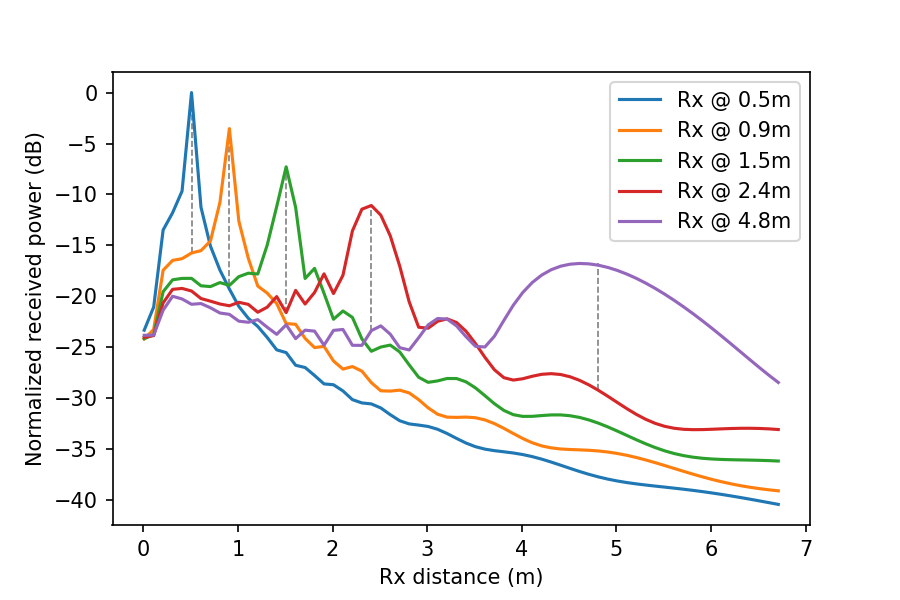}
    \caption{Near-field multiple-access for users with same angular orientation.}
    \label{fig:user_sch}
\end{figure}

In our simulation, we apply Algorithm \ref{alg:1} to $K:=\abs{\sU}=100$ locations, given by $\vr=(0.0,d)$ with $d=0.1m,\dots,10m$ (distributed uniformly on the Z-axis), and set the SIR threshold to $\gamma = 18$dB.
The algorithm outputs the set $\sV$ with $5$ Rx positions that satisfy the SIR threshold.
In other words, at these $5$ positions out of all in $\sU$, the users can be served with a minimum (pairwise) SIR of $18$dB.
In Fig.~\ref{fig:user_sch}, we plot the normalized received power profiles corresponding to each user location in $\sV$, normalized such that the power received at the position closest to the Tx in $\sV$ is $0$dB.
The plot shows that at smaller distances from the Tx, more users can be served, and as the Rx moves farther, it becomes harder to focus the signal at a specific position.

\section{Conclusion and Future Research}
This paper highlights the relevance of near-field models in the context of wireless communication systems with the advent of higher frequencies, larger apertures, and a growing interest in indoor applications. The importance of starting from the precise model, i.e., Maxwell's equations, and establishing an analogy between near-field and far-field models is emphasized. This approach allows for a better understanding of the accuracy-complexity tradeoff, which is crucial in designing efficient communication systems. The paper also illustrates the extent to which approximations are accurate, providing insights into the areas where these models can be reliably applied. Furthermore, two practical applications of near-field models are explored: beamforming and multiple access. In particular, simultaneously serving multiple users in the same direction is demonstrated. For future work, we will address the NLOS channel modeling in the near-field regime.

\bibliography{references}
\bibliographystyle{ieeetr}
\end{document}